\documentclass[
preprint,
]{aastex631}
\makeatletter
\let\frontmatter@title@above\relax
\makeatother
\usepackage{enumitem}
\usepackage{footmisc}
\usepackage[version=4]{mhchem}
\newcommand{\kms}{$\mathrm{km\,s^{-1}}$}

\received{2026 November 23}
\submitjournal{ApJ}

\begin{document}

\title{Laboratory Rotational Spectroscopy and Astronomical Confirmation of C$_5$H$^+$}

\author{H. Gupta}
\email{harshal.gupta@cfa.harvard.edu}
\affiliation{Center for Astrophysics $\vert$ Harvard \& Smithsonian, Cambridge, MA 02138, USA}
\affiliation{National Science Foundation, Alexandria, VA 22314, USA}

\author{P. B. Changala}
\email{changala@jila.colorado.edu}
\affiliation{Center for Astrophysics $\vert$ Harvard \& Smithsonian, Cambridge, MA 02138, USA}
\affiliation{JILA, University of Colorado Boulder and National Institute of Standards and Technology, Boulder, CO 80309, USA}
\affiliation{Department of Physics, University of Colorado Boulder, Boulder, CO 80309, USA}


\author{M. C. McCarthy}
\email{mmccarthy@cfa.harvard.edu}
\affiliation{Center for Astrophysics $\vert$ Harvard \& Smithsonian, Cambridge, MA 02138, USA}

\begin{abstract}
\noindent

The rotational spectrum of the carbon chain cation C$_5$H$^+$ has been measured at high resolution by Fourier transform microwave spectroscopy of a supersonic molecular beam.  Six transitions between 9 and 39\,GHz, measured to better than 0.2\,ppm, enable the rotational and centrifugal distortion constants of C$_5$H$^+$ to be determined to high precision. Five transitions of C$_5$D$^+$ detected at the expected isotopic shift confirm the laboratory assignment. The spectroscopic constants of C$_5$H$^+$ agree precisely with those derived from four sharp rotational emission lines detected previously toward the cold dark cloud TMC-1, thus confirming the astronomical identification of C$_5$H$^+$.  The production of C$_5$H$^+$  using different hydrocarbon precursors, its abundance relative to the smaller carbon chain cation C$_3$H$^+$, and the prospects for the detection of larger carbon chain cations in the laboratory are also discussed.

\end{abstract}


\section{Introduction}

The discovery of new molecular ions in cold dark clouds provides a sensitive probe of the chemical pathways that assemble carbon-bearing molecules under the low-temperature, low-density conditions of the interstellar medium \citep[e.g.,][]{mccall:1941,herbst:1,mcguire:402}. Among molecular clouds, TMC-1 has proven to be an exceptionally rich and observationally favorable source for the detection of new molecules, and therefore an ideal laboratory for testing models of carbon-chain, hydrocarbon, and most recently, aromatic chemistry \citep{mccarthy:345}. The recent astronomical assignment of $l$-\ce{C5H+} (hereafter simply \ce{C5H+}) in TMC-1 \citep{cernicharo:l16} is therefore significant not simply as the addition of another molecule to the interstellar inventory, but as evidence that protonated carbon chains may be more abundant, chemically persistent, and observationally accessible than previously thought. Combined with the detection of $l$-\ce{C3H+} (hereafter simply \ce{C3H+}) in the same source, \ce{C5H+} provides a new benchmark for understanding the balance among ion-neutral formation, destruction by H$_2$, dissociative recombination, and the growth of carbon-chain ions in these extreme environments, where chains as long as \ce{C8H} \citep{mccarthy:l31,mccarthy:158} and \ce{C10H-} \citep{remijan:l45} have been detected.

The chemical significance of \ce{C5H+} lies in the unexpected behavior of the \ce{C3H+}/\ce{C5H+} pair. By analogy with other carbon-chain families, where abundances routinely decrease with increasing molecular size \citep{loomis:188}, one might expect \ce{C5H+} to be less abundant than \ce{C3H+}. Instead, the derived column densities in TMC-1 imply a \ce{C5H+}/\ce{C3H+} ratio of approximately 3.7$\pm$0.5 \citep{cernicharo:l16}, meaning that the larger ion is nearly four times more abundant than \ce{C3H+}.  This is notable because the corresponding neutral linear radical, \ce{C5H}, has a column density of $1.3\pm0.3\times 10^{12}$ cm$^{-2}$ \citep{cabezas:L2}, comparable to published values for \ce{C3H} [$1.1\times 10^{12}$ cm$^{-2}$; \cite{loison:4075}]. Chemical modeling suggests that the elevated column density of \ce{C5H+} relative to \ce{C3H+} reflects differences in destruction chemistry: \ce{C3H+} reacts efficiently with \ce{H2}, whereas \ce{C5H+} is less rapidly removed by this pathway and can therefore persist at a higher steady-state abundance. The \ce{C3H+}/\ce{C5H+} pair thus provides a particularly useful diagnostic of ion-neutral chemistry, molecular growth, and the survival of charged carbon chains under cold-cloud conditions.

Laboratory confirmation is nevertheless essential for precisely this reason. The astronomical assignment was based on a highly regular harmonic progression of lines, characteristic of a closed-shell, linear molecule; close agreement with theoretical predictions among plausible carriers; and chemical arguments informed by the elemental composition and molecular inventory already established in TMC-1 \citep{xue:9}. Even when such an assignment is persuasive, astronomical spectra alone may leave residual ambiguity because alternative carriers cannot always be excluded solely from frequency agreements and source chemistry. Laboratory spectroscopy provides additional experimental constraints on the possible carrier: variation of precursor gases and discharge conditions, isotopic substitution, and other diagnostic tests can be systematically performed to establish the identity of the carrier beyond any reasonable doubt, characterize its formation and relative abundance, and  inform broader questions about the chemistry and detectability of related interstellar ions.

For \ce{C3H+}, the unambiguous identification of the astronomical carrier was ultimately established by \cite{brunken:l4}, who used cryogenic ion-trap spectroscopy and mass-selective action detection to show that the lines arise from \ce{C3H+} and no other molecule. Fourier-transform microwave (FTMW) measurements \citep{mccarthy:10} provided complementary information that is difficult to extract from astronomical data alone. In particular, the laboratory work showed that the fractional abundance of \ce{C3H+} relative to the neutral radical \ce{C3H} is of order $10^{-4}$, toward the lower end of the  $10^{-3}-10^{-4}$ ratios previously found for protonated ions using the same discharge nozzle \citep{gottlieb:1910}. More broadly, this result suggested that positive ions of abundant neutral radicals may be present at experimentally detectable levels, and that new ions may be found either when sufficiently accurate theoretical estimates of their rotational constants are available or, as in the present case, when precise astronomical frequencies provide the likely carrier to target in the laboratory.    

Motivated by the astronomical assignment of \ce{C5H+} in TMC-1, we have undertaken FTMW experiments to determine whether the carrier of the astronomical lines can be reproduced in the laboratory under conditions known to produce transient hydrocarbon ions. Using several hydrocarbon precursors in a supersonic discharge source, a series of rotational transitions whose frequencies agree precisely with those derived from the astronomical spectrum has been observed. In addition, isotopic experiments were subsequently performed using a deuterium-enriched discharge, and an analogous series of lines detected at the frequencies expected for \ce{C5D+}. Finally, the ion-to-neutral ratio was derived and compared with that found for \ce{C3H+}/\ce{C3H}, providing a measure of the extent to which both ions react with \ce{H2} and informing the prospects for detecting \ce{C7H+} and other carbon-chain ions in the laboratory.

\section{Laboratory Measurements}

The laboratory measurements were performed using a cavity FTMW spectrometer coupled to a pulsed supersonic discharge source \citep{mccarthy:611}. This instrument operates from approximately 5 to 40\,GHz and has previously been used to detect numerous transient species, high-energy isomers \citep{mladenovic:174308}, and hydrocarbon ions \citep{mccarthy:l141}, including \ce{C3H+} \citep{mccarthy:10}.  Because \ce{C3H+} is relatively light, only a single transition (at 22.49\,GHz) of this ion lies within the spectrometer band. Even so, that detection provides direct precedent for the present work provided the astronomical carrier is \ce{C5H+}, the next member of the homologous series.   Its larger mass yields a smaller rotational constant and a more closely spaced spectrum, with transitions separated by roughly 4.8\,GHz.  The four astronomical lines assigned to \ce{C5H+} in TMC-1 are the $J=7-6$, $8-7$, $9-8$, and $10-9$ transitions between 33 and 49\,GHz. The lowest two of these transitions fall within the FTMW band and can be compared directly with laboratory measurements,  while all low-$J$ transitions except the fundamental rotational transition are also accessible with present instrumentation and can be predicted with uncertainties of only $\sim$5~kHz. The laboratory investigation therefore required no search in frequency space and focused instead on identifying source and discharge conditions that efficiently produced the carrier.


The source conditions were initially optimized using \ce{C3H+} as a chemical proxy. Previously, the fundamental transition of \ce{C3H+} was quite challenging to detect, requiring nearly 8~hours of total integration to achieve a signal-to-noise ratio (SNR) of 5 \citep{mccarthy:10}.  In the present experiment, however, the line was detected with an
 SNR of 8 after only 14 minutes of integration.  The strongest signals were obtained with a 1000~V (${\sim}30$~mA) dc discharge through diacetylene (\ce{HC4H}) heavily diluted in He ($<0.05\%$). The gas mixture was held at a stagnation pressure of 2.5~kTorr behind the pulsed nozzle. The discharge was synchronized with a $600$\,$\mu$s gas pulse at a repetition rate of 6~Hz, corresponding to a flow of approximately 25-35~SCCM.  As previously observed for \ce{C3H+} \citep{mccarthy:10}, signals of \ce{C5H+} were only observed when helium alone was used as the buffer gas.  Molecular hydrogen, which is commonly added when generating protonated ions---where it serves as the proton donor through \ce{H3+}---was also added to ascertain its effect on the production of \ce{C5H+}.  No enhancement of \ce{C5H+} signals was observed, but nor did the signals  disappear, suggesting that \ce{H2} addition does not appreciably promote reactive loss of the ion, in contrast to what was previously observed for \ce{C3H+} \citep{mccarthy:10}.  Among the noble gases, it is found empirically that He yields the strongest cation signals in this source, the yield declining as
heavier gases such as Ne or Ar are substituted \citep{mccarthy:71}. This latter trend has been attributed to the higher expansion velocity of the lighter gas, which shortens the residence time of the ions and thereby limits their loss through dissociative recombination with electrons.

Under these conditions, a total of six rotational lines were clearly detected between 9 and 39\,GHz, including the astronomical line at 38.6\,GHz.  A seventh line near 33.8\,GHz falls within 0.5\,MHz of a strong $b$-type transition of the \ce{C5H2} ring-chain isomer [33767.463(2)\,MHz; \cite{travers:l135}], so only its low-frequency Doppler component could be detected allowing only an estimate of the rest frequency on the assumption that the Doppler doubling of the line is the same as that of the $c$-\ce{C5H2} line.  
The same lines were subsequently detected using acetylene (\ce{HCCH}) as the hydrocarbon precursor, although intensities were approximately two to three times weaker than those observed with \ce{HC4H}. This precursor dependence is consistent with the greater efficiency in producing longer chain-bearing hydrocarbon species, including \ce{C5H}, when \ce{HC4H} is used instead of \ce{HCCH}. Several experimental diagnostic tests were performed on these lines. They disappeared when the discharge was turned off, demonstrating that the carrier is a transient species produced in the discharge. In addition, the intensities of the lines decreased when a strong permanent magnet was placed near the molecular beam, an expected consequence of deflecting Lorentz forces experienced by charged molecules. 

Differences in the velocity distributions of molecular ions and neutral molecules were investigated by measuring three pairs of rotational transitions of \ce{C5H+} and the similarly sized, closed-shell molecule \ce{H2C5}, spanning 13.7--24.2~GHz. To minimize frequency-dependent instrumental effects, the transitions within each pair were chosen to be separated by less than 1.2~GHz and to share the same upper-state rotational quantum numbers ($J=3,4,5$). The two transitions in each pair were measured under the same discharge conditions. The velocity linewidths and Doppler separations were determined by nonlinear least-squares fitting of the observed profiles with two Lorentzian components of equal width.  The mean ratio of the ionic to neutral velocity linewidths was determined to be $1.34\pm0.04$, where the uncertainty represents the standard deviation across the three transition pairs. The separation between the red- and blue-shifted Doppler components was also found to be larger for \ce{C5H+} than for \ce{H2C5} in these measurements, suggesting that the ions possess axial drift velocities a few percent greater than those of the neutral molecules under the present experimental conditions. The observed enhancements range from approximately 1.5\% to 4.5\%, although their variation exceeds that expected from the formal fitting uncertainties alone. Figure~\ref{fig:profiles} shows one of these transition pairs near 19~GHz.

To remove any possible lingering uncertainty as to the carrier, a search for \ce{C5D+} was performed using a deuterium-enriched discharge of DCCD and He.  As in the original search, the discharge conditions were first optimized on the neutral radical \ce{C5D}, which served as a convenient proxy for the ion. An analogous series of rotational lines was ultimately detected at the frequencies predicted for \ce{C5D+} with comparable signal-to-noise ratio to those of \ce{C5H+} produced with acetylene. The lines disappeared when DCCD was replaced with HCCH, indicating the carrier contained deuterium rather than hydrogen.   The carrier also exhibits the same behavior as \ce{C5H+}: the lines disappeared without a discharge and  diminished in intensity in the presence of a strong magnet.

Finally, the abundance of \ce{C5H+} relative to its parent radical, \ce{C5H}, was estimated to gauge how efficiently the ion is produced. Peak line intensities of both species, recorded under identical conditions, were placed on a common scale by correcting for differences in dipole moments, rotational partition functions, and transition line strengths. The same procedure was applied to \ce{C3H+} and \ce{C3H} for comparison.  Ionic number densities were estimated by comparing line intensities with those of \ce{^{18}OCS} in a calibrated sample of $2\%$ OCS in He, using a procedure similar to that described in earlier works \citep{BalleFlygare1981,gottlieb:1910}, and are good to within a factor of 5.  Table~\ref{tab:densities} summarizes the ion-to-neutral ratios and ionic number densities.

    

\section{Results and Discussion}\label{sec:results}

\begin{figure*}[h]
\centering 
\includegraphics[width=0.7\textwidth]{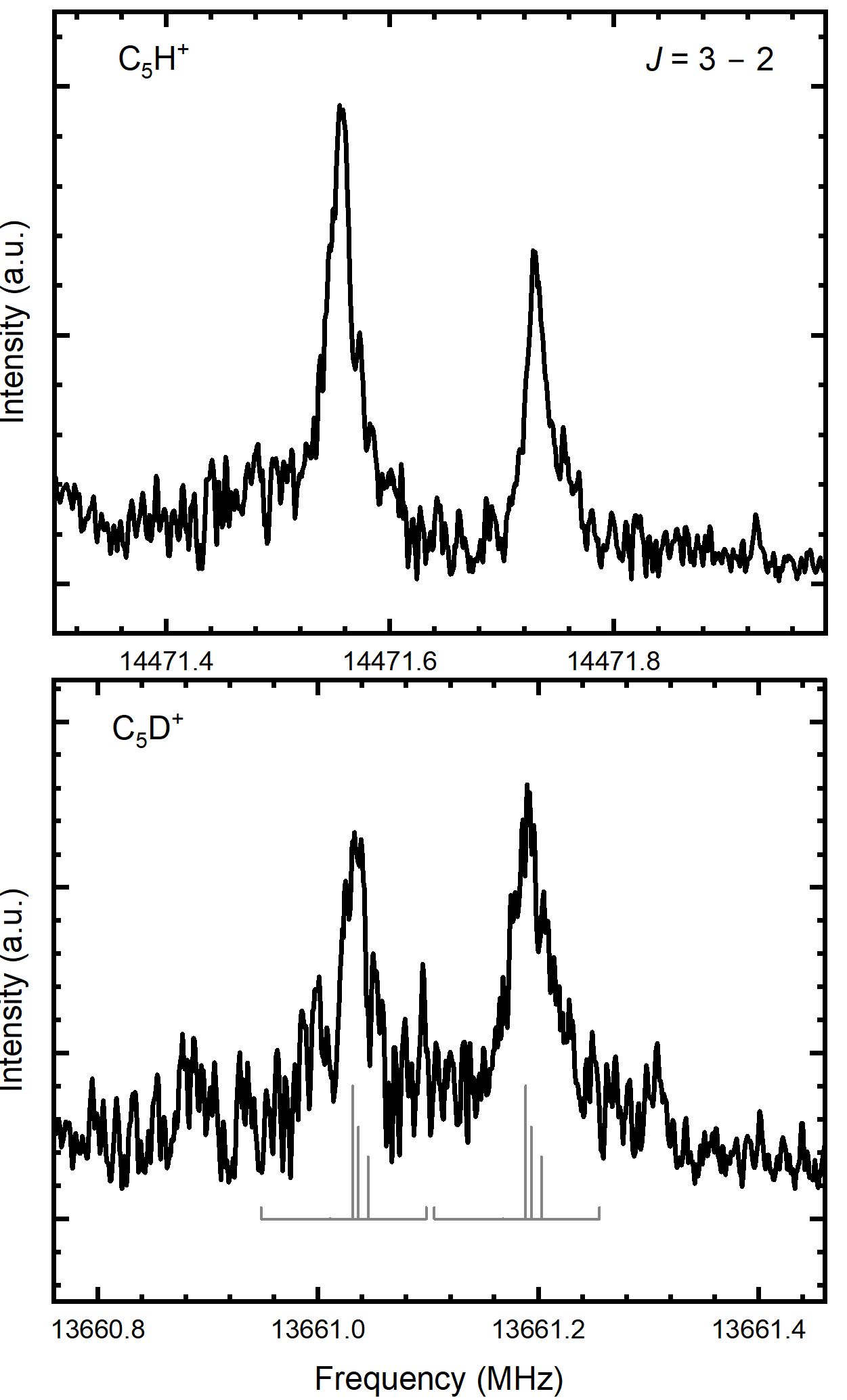}
\caption{\label{fig:sample} Rotational transitions of \ce{C5H+} and \ce{C5D+}.  
The double-peaked Doppler profiles result from the expansion of the molecular beam and its subsequent interaction with the two traveling waves composing the confocal mode of the Fabry--P{\'e}rot cavity.  The \ce{C5H+} spectrum was acquired in about 6 minutes and the \ce{C5D+} spectrum in about 158 minutes of integration.  The stick spectra in gray show the compact deuterium hyperfine structure.
}
\end{figure*}

\begin{figure*}[h]
\centering 
\includegraphics[width=0.5\textwidth]{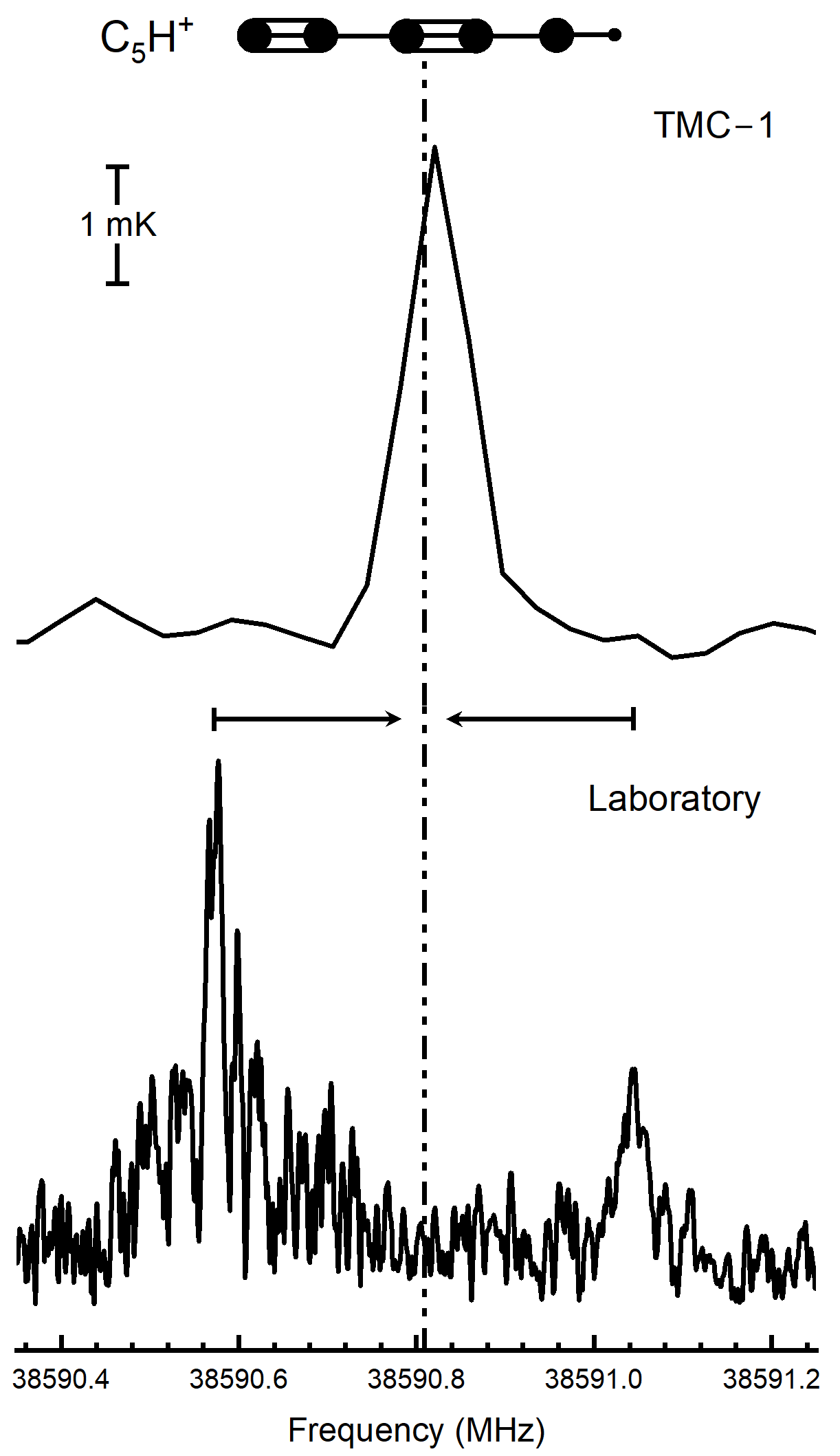}
\caption{\label{fig:lab_vs_astro}
The $J = 8\text{--}7$ line near 38.6\,GHz, as measured in the laboratory by
FTMW spectroscopy (bottom) and toward TMC-1 (top; \citealt{cernicharo:l16}).
The astronomical line center agrees to \textbf{within about 10~kHz---roughly
one-fourth spectral channel (38\,kHz) of the astronomical survey---}with the rest frequency
derived from the laboratory data, as indicated by the dashed vertical line.
The laboratory profile shows the characteristic double-peaked Doppler
structure that arises when the supersonic beam expands along the axis of the
Fabry--P\'erot cavity; the rest frequency corresponds to the mean
of the frequencies of the two Doppler components, whereas the astronomical
frequency reflects the radial velocity of TMC-1 (5.83\,\kms). The laboratory spectrum was acquired in about 4.5 hr of integration and the astronomical spectrum in 1509 hr of integration.}  
\end{figure*}

\begin{figure*}[h]
\centering 
\includegraphics[width=0.7\textwidth]{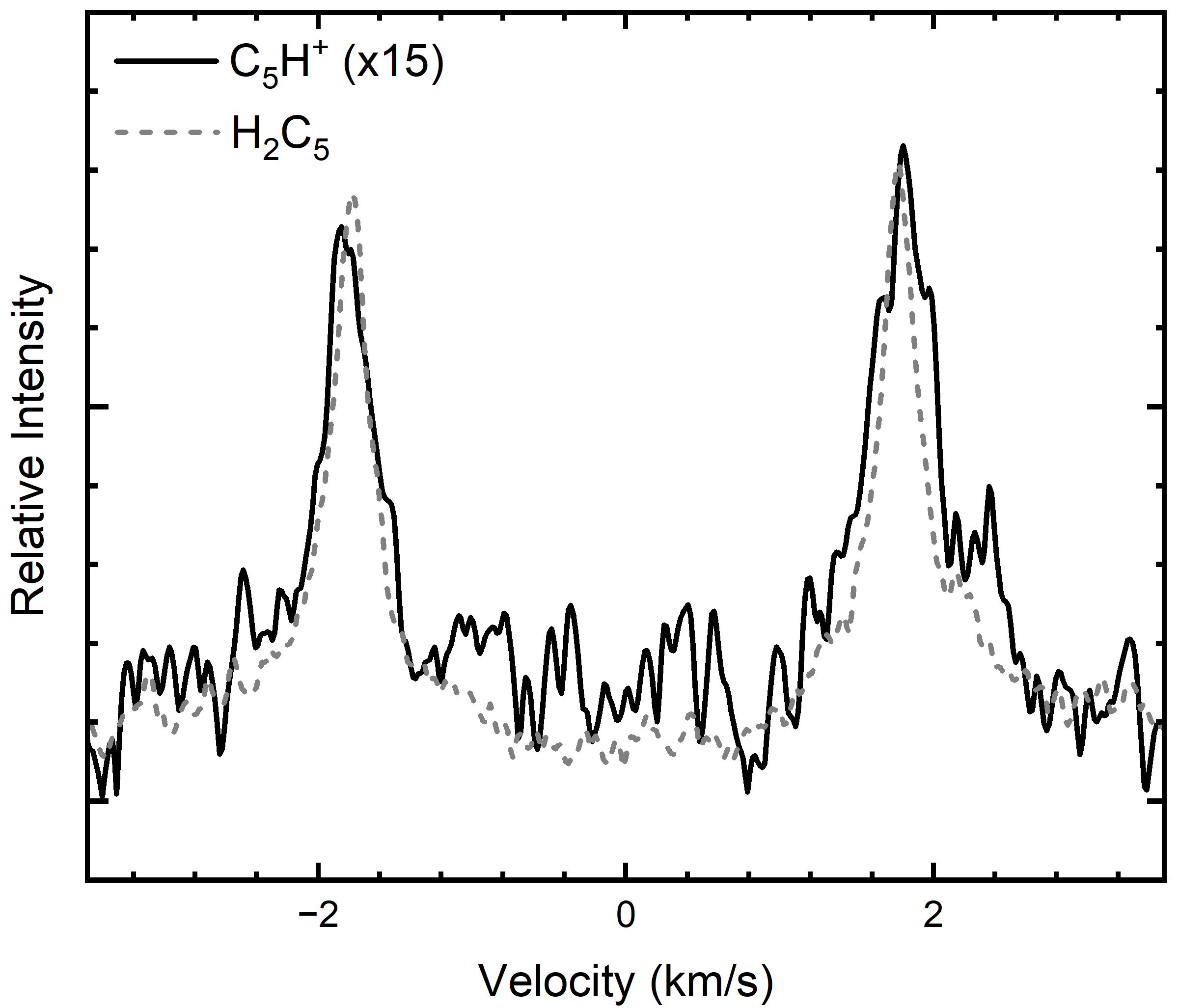}
\caption{\label{fig:profiles} Rotational line profiles of C$_5$H$^+$ compared to those of \ce{H2C5}, a similarly sized neutral molecule observed in the same supersonic beam by FTMW spectroscopy, plotted in velocity. Shown here are the $J=4-3$ transition of C$_5$H$^+$ at 19295.511~MHz and the $J_{K_{a}K_{c}}=4_{04}-3_{03}$ transition of \ce{H2C5} at 18362.323~MHz acquired in about 120 minutes and 33 minutes of integration, respectively.  The spectra have been vertically scaled and offset for clarity.
As in Figure \ref{fig:sample}, the double-peaked Doppler profiles result from the expansion of the molecular beam and its subsequent interaction with the two traveling waves composing the confocal mode of the Fabry--P{\'e}rot cavity. 
}
\end{figure*}

Representative transitions of \ce{C5H+} and its deuterated isotopologue
\ce{C5D+} are shown in Figure~\ref{fig:sample}. The six rotational transitions of \ce{C5H+} measured in the laboratory (Table~\ref{tab:lab_data}) and the four detected toward TMC-1 by \cite{cernicharo:l16} span complementary but only slightly overlapping frequency ranges: the astronomical lines lie mostly above the FTMW band, with only the $J$ = 8–7 transition at 38.6\,GHz common to both data sets. As illustrated in Figure~\ref{fig:lab_vs_astro}, the laboratory and astronomical rest frequencies of this shared transition differ by only 10\,kHz, comparable to the combined $1\sigma$ uncertainty
of the two measurements, noting that the astronomical value must be transformed to the rest frame using the systemic (LSR) velocity of TMC-1, a correction that carries a small additional uncertainty of its own. The agreement of this single shared line already points strongly to a common carrier.

A comparison of the derived molecular parameters (Table~\ref{tab:constants})
makes the same point with even greater force. The laboratory and astronomical
rotational constants differ by less than 1\,kHz---well within the $1\sigma$
uncertainty of the astronomical value, which is itself roughly three times
larger than that of the laboratory determination. The centrifugal distortion
constants, $D = 136.6(24)$\,Hz and $138(3)$\,Hz, respectively, are likewise
indistinguishable, differing by less than the $1\sigma$ uncertainty derived from either of the two data sets. Given that the two determinations rest on wholly
independent data sets measured over different frequency ranges, this
near-perfect agreement of the spectroscopic constants constitutes definitive
confirmation that the same molecule has been observed both in the laboratory
and in space.  Two further properties follow directly from the laboratory data: the absence of any line structure even in the low-$J$ transitions, where hyperfine splittings would be largest, and at the high resolution achieved here ($<0.2$\,ppm) confirms the carrier is a closed-shell molecule, and its production from \ce{HC4H} and \ce{HCCH} precursors alone indicates it is a hydrocarbon. This at once excludes the nitrogen- and oxygen-bearing carriers with comparable rotational constants considered by \citet{cernicharo:l16}, such as \ce{C4N}, \ce{C4O}, and \ce{HC4O}, independently of the spectroscopic grounds on which they were already disfavored.   Among hydrocarbons, the only candidates with approximately the correct rotational constant are linear \ce{C5H} in its neutral, cationic, and anionic forms.

Of these three forms, the neutral radical can be excluded immediately: the rotational spectrum of \textit{l}-\ce{C5H} is well characterized at low frequency and high spectral resolution \citep{mccarthy:174308, mccarthy:158}, its rotational constant differs  by 0.7\% from the value measured here, and, as an open-shell $^{2}\Pi$ radical, it displays fine and hyperfine structure, whereas the transitions observed here show none. The anion can be ruled out as well, on essentially the same grounds established in the original astronomical assignment but reinforced here by new higher-level calculations and higher resolution spectra. In their original assignment, \citet{cernicharo:l16} identified \ce{C5H-} as the only other viable alternative but excluded it for three reasons: the rotational constant derived from the astronomical spectrum agrees to within 0.07\% with a scaled theoretical value for \ce{C5H+}, whereas it differs by 2.7\% from the value extracted from the contour of an electronic band of the linear triplet anion ($B_0 = 2476.3\pm6$\,MHz; \citealt{tulej:6878}); no hyperfine structure is present in the astronomical lines; and the anion is not detected toward the carbon-rich envelope IRC+10216, where the other anions known in TMC-1 appear. Because the present measurements are consistent with ionic \ce{C5H} as the carrier, that reasoning applies directly here.
The higher-level coupled-cluster calculations presented in Appendix~\ref{sec:calc} strengthen the first of these arguments: the best-estimate purely \textit{ab initio} rotational constant of \ce{C5H+} (2410.136\,MHz) agrees with the measured value to within 0.07\% even without empirical scaling---the same level of agreement that \citet{cernicharo:l16} obtained only after scaling---and to within 0.02\% (2411.409\,MHz) once scaled by the predicted-to-measured ratio of \ce{C3H+}.   Reproducing the observed spectrum with \ce{C5H-} would thus require a closed-shell, linear anion, and no such species is supported by either theory or experiment. Calculations find several low-lying electronic states close in energy but disagree on their ordering: \citet{bennedjai:255} place the global minimum at a cyclic three-membered ring and question whether a linear form is a true minimum at all, whereas \citet{satpati:115} appear to favor a low-lying linear chain with a $^{3}\Sigma^{-}$ ground state. A cyclic anion is a near-prolate asymmetric top
 whose rotational constants bear no resemblance to those measured here, whereas a linear anion is, according to both anion photoelectron spectroscopy
\citep{sheehan:174301} and detachment spectroscopy \citep{tulej:865}, an open-shell triplet chain. Being open-shell, the anion would  display fine structure from the unpaired electron spins together with proton hyperfine splitting. Yet at the roughly threefold finer resolution achieved in the laboratory compared to the TMC-1 observations (15\,kHz vs. 38\, kHz), the transitions detected here remain single, featureless lines, showing none of the fine or hyperfine structure an open-shell anion would exhibit.

Deuterium isotopic substitution provides a further, independent line of
evidence, and one that decisively identifies the carrier as
\ce{C5H+} and no other molecule. Using the same least-squares analysis, the
rotational and centrifugal distortion constants $B$ and $D$ were fit to the
five measured transitions of \ce{C5D+}. The rms of the fit, 2.0~kHz, is
comparable to that obtained for \ce{C5H+} (1.4~kHz; Table~\ref{tab:constants}),
and both constants are determined with comparable precision---$B$ to better
than 1~part in $10^{6}$---and agree very closely with the values scaled from
the parent species (Appendix~\ref{sec:calc}). The magnitude of the isotopic
shift is exactly that expected for substitution of D for H at the terminus of
a linear \ce{C5H+} backbone. Unlike hydrogen, the deuterium nucleus possesses
an electric quadrupole moment, but its coupling constant, $eQq$, is small
[e.g., $\sim$0.2\,MHz for \ce{C5D}; \cite{hirota:196}]. Given the wider
linewidths characteristic of a fast-moving He beam, the resulting hyperfine
structure in \ce{C5D+} is expected to be at most partially resolved and only in the lowest
few $J$ transitions; the fundamental $J = 1\text{--}0$ line, where the
splitting is largest, lies below the frequency range of the spectrometer. In
the lowest transition we observe, $J = 3\text{--}2$, the unresolved quadrupole
structure should slightly broaden the line as the calculated stick spectrum of
Figure~\ref{fig:sample} indicates. Although the corresponding D-isotopic shift of the neutral radical
\ce{C5H} is very similar (5.88\% vs 5.92\%), that species is excluded above by
its differing rotational constant and its open-shell fine and hyperfine
structure.


The laboratory line profiles provide a further consistency check and evidence that the carrier is an ion rather than a neutral molecule. As shown in Figure~\ref{fig:profiles}, the full widths at half maximum (FWHM) of the \ce{C5H+} lines are substantially larger than those of \ce{H2C5}, with a mean enhancement of $34\%$ across the three transition pairs measured. The linewidth ratios are consistent within their uncertainties, although the frequency dependence of the instrumental response has not been fully characterized. Such enhanced linewidths have previously been reported for positively charged species produced in this supersonic discharge source \citep{gottlieb:1910}. The larger collisional cross sections associated with long-range ion--neutral interactions may contribute to the observed broadening, although other discharge-related and instrumental effects cannot be excluded. Under the discharge conditions employed for these measurements, the separation between the red- and blue-shifted Doppler components is also larger for \ce{C5H+}, corresponding to axial drift velocities roughly $1.5$--$4.5\%$ greater than those of \ce{H2C5}. This difference is observed to depend on the discharge conditions, and therefore the larger separation itself is not interpreted as a general characteristic of molecular ions. The broader \ce{C5H+} line profiles, the observed differences in axial drift velocities, and the reduction in line intensity when a permanent magnet is brought near the molecular beam, consistent with the magnetic deflection of charged particles, provide compelling support for the ionic character of the carrier. 

Taken in totality, this evidence---the precise agreement of the shared transition and of the derived spectroscopic constants with the astronomical data; the closed-shell hydrocarbon character constrained by the discharge chemistry and the absence of any line structure; the corroborating deuterium substitution; and the ionic character revealed by the line profiles, drift velocities, and magnetic deflection---establishes the assignment of the carrier to \ce{C5H+} beyond any reasonable doubt.

Beyond confirming the identification, the laboratory measurements constrain
the production chemistry of \ce{C5H+} and its relationship to the smaller
homolog \ce{C3H+}. Table~\ref{tab:densities} lists the inferred number
densities per gas pulse and the corresponding ion-to-neutral ratios for both
ions with the HCCH and \ce{HC4H} precursors. The most striking result is that the
ion-to-neutral ratio is far more favorable for the \ce{C5H} system:
$n(\ce{C5H+})/n(\ce{C5H}) \approx 4-7\times10^{-3}$, nearly independent of the two
precursors, exceeds the corresponding \ce{C3H+}/\ce{C3H} ratio by
one to two orders of magnitude. In absolute terms as well, \ce{C5H+} is
produced at least as abundantly as \ce{C3H+} when acetylene is used as the precursor ($240\times10^{8}$ versus $80\times10^{8}$ molecules per pulse), and roughly 17 times more abundantly when diacetylene is used as the precursor
($730\times10^{8}$ versus $45\times10^{8}$ molecules per pulse). The larger ion thus
maintains a higher steady-state abundance than the smaller one under
laboratory conditions---the same counterintuitive ordering found in TMC-1,
where the \ce{C5H+}/\ce{C3H+} column-density ratio is
$3.7\pm0.5$ \citep{cernicharo:l16}. That the laboratory discharge reproduces
this behavior lends direct experimental support to the chemical picture in
which \ce{C5H+}, unlike \ce{C3H+}, is either not or at the very least only slowly destroyed by reaction with \ce{H2} and therefore persists at an elevated fractional abundance.

It is plausible that the formation routes for \ce{C5H+} and \ce{C3H+} in our laboratory discharge are different from ones operative in cold dark clouds like TMC-1.  While protonation of \ce{C5} and \ce{C3} by \ce{H3+} is modeled as the predominant route for the production of \ce{C5H+} and \ce{C3H+} in TMC-1 \citep{cernicharo:l16}, the observed abundances and abundance ratios in the laboratory suggest that this is unlikely to be the case in our pulsed discharge nozzle.  Flowing afterglow measurements show that \ce{C+} insertion into \ce{HC4H} yields \ce{C5H+} and \ce{C3H+} with product branching ratios of 0.5 and 0.05, respectively, or an abundance ratio of about 10:1 \citep{Dheandhanoo1986}.   The observed \ce{C5H+/C3H+} abundance ratio in our \ce{HC4H}/He discharge is about 17:1, remarkably similar to that found in those experiments, which perhaps represent the closest comparison available for the high pressure conditions prevalent in the throat of the supersonic nozzle used in our experiment.  Similarly, the \ce{C3H+} so produced may undergo additional reactions with \ce{HC4H} to yield \ce{C5H+}.  Although similar formation routes of \ce{C5H+} via \ce{C+} insertion into HCCH are not known, it is plausible that \ce{HC4H} is produced at sufficiently high abundances in an HCCH discharge \citep{Vasile1977} to yield \ce{C5H+} via \ce{C+} insertion or reactions with \ce{C3H+}, thereby explaining the roughly equal observed abundances of \ce{C5H+} and \ce{C3H+} in our HCCH/He discharge.


These results have encouraging implications for the detection of other carbon-chain cations. For example, the favorable ion-to-neutral ratio measured for the \ce{C5H} system, combined with the sensitivity of the FTMW technique, places the next member of the series, \ce{C7H+}, within experimental reach, even though the steady-state concentration of \ce{C7H} is only about one-sixth that of \ce{C5H}. If \ce{C7H+}, like \ce{C5H+}, is either unreactive toward \ce{H2} or only slowly destroyed by it, its ion-to-neutral ratio should be comparably favorable, and the modest integration times required here for \ce{C5H+} would largely offset the decline in neutral abundance with molecular size. Its smaller rotational constant, moreover, means that multiple transitions lie within the frequency range of the spectrometer, so that a targeted search guided by accurate ab initio constants would require limited exploration in frequency space, and any detection could be confirmed readily. This strategy has, in fact, already proven effective beyond hydrocarbon cations: by the same method, we have recently detected the open-shell radical cation \ce{HC3N+} in a dilute cyanoacetylene (\ce{HC3N}) discharge \citep{gupta:hc3n}, confirming its astronomical assignment in TMC-1 \citep{cabezas:l22}. The confirmation of \ce{C5H+} reported here, together with that of \ce{HC3N+}, therefore not only secures the identity of new interstellar ions but also charts a practical route toward the laboratory and astronomical detection of a wider range of molecular cations that plausibly persist in cold dark clouds.
\pagebreak

\section*{Acknowledgments}

We thank J. Cernicharo for helpful discussions, and sharing the data required to make Figure~\ref{fig:lab_vs_astro}. P.B.C. and M.C.M. are supported by the National Science Foundation (award nos.  AST-1908576 and PHY-2110489).  Additional contributions by P.B.C. at JILA were supported by NIST.  H.G. acknowledges support from the National Science Foundation for participation in this work as part of his independent research and development plan.  Any opinions, findings, and conclusions expressed in this material are those of the authors and do not necessarily reflect the views of the National Science Foundation.

This document has not been peer reviewed but has been cleared by NIST for release.

\pagebreak

\appendix

\section{Electronic Structure calculations}\label{sec:calc}
The equilibrium geometry of \ce{C5H+} was optimized with coupled cluster theory including single, double, and perturbative triple excitations~\citep{Raghavachari1989:CCSDt,Bartlett1990:CCSDt} [CCSD(T)] and analytic gradients~\citep{Lee1991:CCSDtgrad,Scuseria1991:CCSDtgrad} as implemented in the CFOUR software package~\citep{Matthews2020:cfour,CFOUR}. Both frozen-core (fc) and all-electron (ae) calculations were performed with the corresponding valence-only and core-valence correlation-consistent basis sets (cc-pCV$X$Z, $X$ = D,T,Q,5)~\citep{Dunning1989:ccpvxz,Woon1995:ccpcvxz}. High-order-correlation corrections to the geometry were derived from perturbative quadruple excitation treatments [CCSD(T)$_\Lambda$], while zero-point contributions to the ground-state rotational constants were derived with second-order vibrational perturbation theory (VPT2)~\citep{Mills1972:vpt} using cubic force constants derived from numerical differentiation of analytic gradients. The theoretical contributions to the ground-state rotational constant of \ce{C5H+} are summarized in Table~\ref{tab:theory}, where we report a best-estimate \textit{ab initio} value of 2410.136\,MHz, only 0.07\% below the measured value. The same theoretical procedure applied to \ce{C3H+} yields a best-estimate B$_0$ of 11239.010\,MHz compared to the measured value of 11244.950(2)\,MHz~\citep{brunken:l4}. If we scale the purely \textit{ab initio} prediction for \ce{C5H+} by the predicted-to-measured ratio of \ce{C3H+}, we derive a scaled prediction for \ce{C5H+} of 2411.409\,MHz, now only 0.02\% below the measured value.

We also computed the centrifugal distortion constant of \ce{C5H+} to be $D=102$\,Hz using the fc-CCSD(T)/cc-pVQZ harmonic force field, in good agreement with \cite{cernicharo:l16}. Following their same scaling strategy, this parameter is predicted to equal 127\,Hz after scaling by the theory/expt ratio of $D$(\ce{C5H}).

The isotope shift for \ce{C5D+} was calculated using a best-estimate equilibrium geometry derived from the fc-CCSD(T)/cc-pV5Z optimized geometry plus a core-correlation correction with a quadruple-zeta basis set and a valence-only CCSD(T)$_\Lambda$ correction derived with a double-zeta basis set. The calculated ratio of the $B_e$ values is $0.943977$. Applying this correction to the measured $B_0$ value for \ce{C5H+} yields a prediction for \ce{C5D+} of 2276.819\,MHz, in very close agreement with the measured $B_0$ value. We also derive a scaled prediction for $D$(\ce{C5D+}) = 115\,Hz by assuming $D \propto B^3$ (a good approximation for semi-rigid linear molecules) and using the measured $B$(\ce{C5D+}/\ce{C5H+}) ratio.

\pagebreak

\bibliography{references,sorted_entries_new}
\bibliographystyle{aasjournal}

\clearpage 
\pagebreak

\begin{deluxetable}{CRRCRRCRR}
\tablecaption{Laboratory and astronomical rotational frequencies of
C$_5$H$^+$ and laboratory frequencies of C$_5$D$^+$. \label{tab:lab_data}}
\tablewidth{0pt}
\tablehead{
\colhead{} & \multicolumn{2}{c}{C$_5$H$^+$ (lab.)} & &
\multicolumn{2}{c}{C$_5$H$^+$ (TMC-1)\tablenotemark{a}} & &
\multicolumn{2}{c}{C$_5$D$^+$ (lab.)} \\
\cline{2-3} \cline{5-6} \cline{8-9}
\colhead{Transition} & \colhead{Frequency} & \colhead{$O-C$} & &
\colhead{Frequency} & \colhead{$O-C$} & &
\colhead{Frequency} & \colhead{$O-C$} \\
\colhead{$J' - J''$} & \colhead{(MHz)} & \colhead{(kHz)} & &
\colhead{(MHz)} & \colhead{(kHz)} & &
\colhead{(MHz)} & \colhead{(kHz)}
}
\startdata
2-1     &  9647.766(1) & -2.3         & & \cdots        & \cdots  & & \cdots       & \cdots  \\
3-2     & 14471.645(1) &  0.8         & & \cdots        & \cdots  & & 13661.114(3) & -0.5    \\
4-3     & 19295.511(1) &  0.7         & & \cdots        & \cdots  & & 18214.806(3) & -0.1    \\
5-4     & 24119.363(2) & -0.3         & & \cdots        & \cdots  & & 22768.486(3) & -0.4    \\
6-5     & 28943.200(2) &  0.2         & & \cdots        & \cdots  & & 27322.157(5) &  4.4    \\
7-6     &  33767.017(5)\tablenotemark{b}    & 0.3 & & 33767.026(5)  & -0.2    & & 31875.801(3) & -0.8    \\
8-7     & 38590.808(5) & -2.7         & & 38590.818(5)  & -2.8    & & \cdots       & \cdots  \\
9-8     &  \cdots      & \cdots       & & 43414.594(5)   & 5.1 & & \cdots       & \cdots  \\
10-9    &  \cdots      & \cdots       & & 48238.325(5)  & -2.2  & & \cdots       & \cdots  \\
\enddata
\tablecomments{Estimated $1\sigma$ uncertainties are given in parentheses in
units of the last digit. The Observed $-$ Calculated values are derived from the best-fit constants given in Table~\ref{tab:constants}.\\
\tablenotemark{a} \citet{cernicharo:l16}.\\
\tablenotemark{b} Blue Doppler component obscured by the strong interloping red component of the $J_{K_aK_c} = 5_{14}-5_{05}$ transition of c-\ce{C5H2}. Frequency estimated on the assumption that the Doppler splitting of 0.398~MHz for \ce{C5H+} is identical to that of c-\ce{C5H2}.
}
\end{deluxetable}


\pagebreak

\begin{deluxetable}{l C C C C C C}
\tablewidth{0pt}
\tablecaption{Spectroscopic Constants of C$_5$H$^+$ and C$_5$D$^+$\label{tab:constants}}
\tablehead{
\colhead{}  &  \multicolumn{3}{c}{C$_5$H$^+$} & & \multicolumn{2}{c}{C$_5$D$^+$} \\
\cline{2-4}   \cline{6-7}
\colhead{Constant}  &  \colhead{Laboratory}  &  \colhead{Astronomical\tablenotemark{a}}   &  \colhead{Predicted\tablenotemark{b}} & & \colhead{Laboratory}  & \colhead{Predicted\tablenotemark{b}}    
}
\startdata
$B$~(MHz)         &  2411.943156(128)    &  2411.94397(55)   &  2411.4  & &  2276.85454(36) & 2276.819 \\
$10^{6}D$~(MHz)   &  136.62(237)    &  138(3)   &  127  & & 117.9(49)
 & 114 \\
\hline
No.~lines &  7  &   4   & \cdots  & &  5  & \cdots\\
rms~(kHz) &  1.4  & 4.4     &  \cdots & &  2.0  & \cdots \\
\enddata
\tablecomments{Best-fit constants derived from least-squares fits to the rotational transitions listed in Table~\ref{tab:lab_data}. Laboratory constants for C$_5$H$^+$ and C$_5$D$^+$ were determined from the rest frequencies using the CALPGM suite of programs \citep{pickett:1991,drouin:2017}; the corresponding input and output files are provided as supplementary material.}
\tablenotetext{a}{From \citet{cernicharo:l16}.}
\tablenotetext{b}{Scaled rotational and centrifugal distortion constants calculated as described in Appendix \ref{sec:calc}.}
\end{deluxetable}


\pagebreak

\begin{deluxetable}{cCCCC}
    \tablecaption{Number densities and ion-to-neutral abundance ratios of \ce{C5H+} and \ce{C3H+}. \label{tab:densities}}
\tablewidth{0pt}
\tablehead{
\colhead{} & \colhead{$n_\mathrm{C_5H^+}$} & \colhead{$\mathrm{C_5H^+}/\mathrm{C_5H}$}  & \colhead{$n_\mathrm{C_3H^+}$} & \colhead{$\mathrm{C_3H^+}/\mathrm{C_3H}$} \\
\colhead{Precursor/Buffer}    &  \colhead{$\mathrm{(10^{8}/pulse)}$}  &  \colhead{Ratio}  &   \colhead{$\mathrm{(10^{8}/pulse)}$}  &  \colhead{Ratio}
}
\startdata
HCCH/He          & 240 & 6.8\times10^{-3}  & 80     &  3.7\times10^{-5}   \\
HC$_{4}$H/He     & 730 & 3.6\times10^{-3}  & 45     &  1.5\times10^{-4}   \\
\enddata
\end{deluxetable}

\begin{deluxetable}{lD}
\tablecaption{Best-estimate theoretical prediction of $B$(\ce{C5H+}). \label{tab:theory}}
\tablewidth{0pt}
\tablehead{
\colhead{Contribution} & \colhead{Value (MHz)}
}
\startdata
Valence CBS $B_e$\tablenotemark{a} & 2393.254 \\
Core-valence correlation\tablenotemark{b} & +10.974 \\
High-order correlation\tablenotemark{c} & $-$2.209\\
Vibrational correction\tablenotemark{d} & +8.117 \\
Total \textit{ab initio} $B_0$ & 2410.136\\
Scaled $B_0$\tablenotemark{e} & 2411.409
\enddata
\tablecomments{
\tablenotemark{a} Valence-only complete-basis-set estimate derived from an exponential extrapolation of the fc-CCSD(T)/cc-pV$X$Z $B_e$ values ($X$ = T,Q,5).\\
\tablenotemark{b} Core-valence correlation correction derived from the difference of fc-CCSD(T)/cc-pVQZ and ae-CCSD(T)/cc-pCVQZ $B_e$ values.\\ 
\tablenotemark{c} High-order correlation correction derived from the difference of fc-CCSD(T)/cc-pVDZ and fc-CCSDT(Q)$_\Lambda$/cc-pVDZ $B_e$ values. \\ 
\tablenotemark{d} Zero-point vibrational corrections derived from VPT2 using fc-CCSD(T)/cc-pVQZ force constants.\\
\tablenotemark{e} Scaled prediction derived from the theory/measured ratio of \ce{C3H+}.
}
\end{deluxetable}

\pagebreak

\pagebreak

\end{document}